# Compressibility and structural stability of ultra-incompressible bimetallic interstitial carbides and nitrides

D. Errandonea[*,†], J. Ruiz-Fuertes[†], J. A. Sans[†], D. Santamaría-Perez[‡], O. Gomis[§], A. Gómez[⊥], and F. Sapiña[⊥]

[†] Departamento de Física Aplicada, ICMUV, Universitat de València, C. Dr. Moliner 50, 46100 Burjassot, Spain

[‡] Departamento de Química-Física I, Universidad Complutense de Madrid, Avda. Complutense s/n, 28040 Madrid, Spain

[§] Centro de Tecnologías Físicas, Universitat Politècnica de València, Camı de Vera s/n, 46022 Valencia, Spain

[⊥] ICMUV, Universitat de Valencia, Apartado de Correos 22085, 46071 Valencia, Spain

**Abstract:** We have investigated by means of high-pressure x-ray diffraction the structural stability of $Pd_2Mo_3N$, $Ni_2Mo_3C_{0.52}N_{0.48}$, $Co_3Mo_3C_{0.62}N_{0.38}$, and $Fe_3Mo_3C$. We have found that they remain stable in their ambient-pressure cubic phase at least up to 48 GPa. All of them have a bulk modulus larger than 330 GPa, being the least compressible material $Fe_3Mo_3C$, $B_0 = 374(3)$ GPa. In addition, apparently a reduction of compressibility is detected as the carbon content increased. The equation of state for each material is determined. A comparison with other refractory materials indicates that interstitial nitrides and carbides behave as ultra-incompressible materials.

[*] FAX: (+34) 96 354 3146. Tel.: (+34) 96 354 4475. E-mail: daniel.errandonea@uv.es

A great effort is currently focused on the synthesis and characterization of materials exhibiting very low compressibility and large durability.[1] In particular, transition metal nitrides and carbides have attracted special attention.[2] These refractory materials have numerous industrial applications and a promising future, on top of being of great interest to the scientific community. Many incompressible nitrides have been recently discovered combining the application of high-pressure (HP) and high-temperature. However, only small amounts of them can be synthesized and their industrial preparation presents several difficulties. In contrast, interstitial bimetallic nitrides can be prepared in large amounts at ambient pressure,[3] having them (e.g. $Ni_2Mo_3N$, $Co_3Mo_3N$, and $Fe_3Mo_3N$) similar mechanical properties than many ultra-hard compounds.[4] In particular, their compressibility is smaller than that of cubic $Si_3N_4$ and comparable to that of cubic BN.[4] Thus, interstitial nitrides could be considered as potential cheap substitutes for now-a-days industrial refractory materials. Isomorphic carbides[5] could be probably even better candidates since the substitution of nitrogen by carbon is expected to reduce compressibility.[6]

In this work we extend the study of the compressibility and structural properties of molybdenum bimetallic compounds to $Pd_2Mo_3N$, $Fe_3Mo_3C$, $Ni_2Mo_3C_{0.52}N_{0.48}$, and $Co_3Mo_3C_{0.62}N_{0.38}$. They were studied by HP x-ray diffraction (XRD) up to 48 GPa, being found to remain in their ambient-pressure cubic structure. They are also ultra-incompressible materials, with $Fe_3Mo_3C$ challenging the incompressibility of cubic BN.

$Pd_2Mo_3N$ was synthesized by ammonolysis of a crystalline precursor following the procedure described by El-Himri *et al*.[3] $Co_3Mo_3C_{0.62}N_{0.38}$, $Ni_2Mo_3C_{0.52}N_{0.48}$, and $Fe_3Mo_3C$ were synthesized by carburization of the corresponding nitrides.[7-9] The nitrides were heated at 10 K/min up to 923 K under flowing of a carburizing mixture of gases (360 $cm^3$/min, 4.4% $CH_4$, 1.1%, 94.5% Ar). The temperature was hold for 2 h (Fe and Co compounds) and 4 h (Ni compound). Then, the solids were quenched to room temperature (RT). The compositions of the compounds were determined by energy-dispersive x-ray analysis on a scanning electron microscope (Phillips XL-30) operating at 20 kV, and equipped with a super ultra-thin detector, which allows the detection of light elements. The carbon and nitrogen content was also evaluated by standard combustion analysis (Carbo Erba Elemental Analizer 1108) with a precision of 0.2%.

Angle-dispersive x-ray diffraction (ADXRD) experiments at RT and HP were carried out at beamline I15 of the Diamond Light Source using a diamond-anvil cell (DAC) and a monochromatic x-ray beam ($\lambda$ = 0.4246 Å). Samples were loaded in a 100-μm hole of a rhenium gasket in a DAC with diamond-culet sizes of 300 μm. Pressure was determined using ruby fluorescence[10]. For $Fe_3Mo_3C$ a second experiment was carried out using gold as pressure marker.[11] Both scales are reliable and allow determining pressure *in situ*. As we will see below, the choice of one pressure scale or the other does not influence the results. A 16:3:1 methanol-ethanol-water mixture was used as pressure-transmitting medium.[12,13] The x-ray beam was focused down to 30 x 30 $\mu m^2$. A pinhole placed before the sample position was used as a clean-up aperture. The images were collected using a MAR345 image plate located at 350 mm from the sample. The diffraction patterns were integrated and corrected for distortions using FIT2D. The structural analysis was performed using POWDERCELL and GSAS.

At ambient pressure, the obtained diffraction patters for $Pd_2Mo_3N$, $Ni_2Mo_3C_{0.52}N_{0.48}$ (space group *$P4_132$*)[5], $Co_3Mo_3C_{0.62}N_{0.38}$, and $Fe_3Mo_3C$ (space group *$Fd\bar{3}m$*, η-carbide structure)[9] correspond to the known cubic structures, which are shown in Fig. 1. The obtained unit-cell parameters are summarized in Table I, showing good agreement with previous data.[3,5,9] Both structures have been already extensively described, being their basic framework formed by polyhedral units with short Mo-N (or Mo-C) bond distances. Fig. 2 shows ADXRD data for $Ni_2Mo_3C_{0.52}N_{0.48}$ and $Fe_3Mo_3C$ at selected pressures. Under compression, the only changes we observed in the XRD patterns are the shift of Bragg peaks toward high 2θ angles and the typical broadening of DAC experiments.[14,15] Similar evolution of diffraction patterns upon compression is found in the other compounds. For $Pd_2Mo_3N$ and $Ni_2Mo_3C_{0.52}N_{0.48}$, all measured XRD patterns can be indexed within the ambient-pressure structure. For $Co_3Mo_3C_{0.62}N_{0.38}$ and $Fe_3Mo_3C$, all measured patterns can be assigned to the η-carbide structure. No evidence of pressure-induced phase transition or chemical decomposition is detected in any of the experiments and the ambient-pressure XRD patterns are fully recovered upon decompression.

From the refinement of the diffraction patterns, we obtained the pressure evolution of unit-cell parameters (and volume) in all compounds. We also refined the internal atomic positions and found that their change with pressure is comparable to experimental uncertainty. Therefore, we decided to neglect the pressure effect on atomic positions. As can be seen in Fig. 2, good fittings are obtained in the refinements. The residuals for $Fe_3Mo_3C$ at 0.9 GPa are $R_{WP}$=2.42%, $R_P$=1.73%, and $R_F^2$=1.51% (55 reflections) and for $Ni_2Mo_3C_{0.52}N_{0.48}$ $R_{WP}$=1.43%, $R_P$=1.91%, and $R_F^2$=1.14% (49 reflections). Comparable values were obtained at different pressures in the four studied compounds. The pressure dependence of the unit-cell volume is plotted in Fig. 3. There it can be seen that the compounds here studied are highly incompressible. The pressure-volume data have been analyzed using a third-order Birch-Murnaghan equation of state (EOS). The obtained volume at ambient pressure ($V_0$), bulk modulus ($B_0$), and its pressure derivative ($B_0$') are summarized in Table I. The bulk modulus of $Ni_2Mo_3C_{0.52}N_{0.48}$ (344 ± 4 GPa) is 4% larger than that of $Ni_2Mo_3N$ (330 ± 8 GPa).[4] The bulk modulus of $Pd_2Mo_3N$ (333 ± 3 GPa) is around 6% larger than theoretical values (312 – 313 GPa)[4,17] and comparable to the bulk modulus of $Ni_2Mo_3N$.[4] The bulk modulus of $Co_3Mo_3C_{0.62}N_{0.38}$ (353 ± 4 GPa) is 1% larger than that of $Co_3Mo_3N$ (350 ± 8 GPa).[4] The bulk modulus of $Fe_3Mo_3C$ (374 ± 3 GPa) is 2% larger than that of $Fe_3Mo_3N$ (368 ± 9 GPa).[4] In particular, $Fe_3Mo_3C$ turns out to be as incompressible as cubic BN. Note that for $Fe_3Mo_3C$ two independent experiments were performed using different pressure scales (gold and ruby) and similar results were obtained (see Fig. 3). According with the above summarized results, apparently substitution of N by C produces an increase of the bulk modulus. This effect is most evident when comparing $Ni_2Mo_3C_{0.52}N_{0.48}$ with $Ni_2Mo_3N$, since in other compounds the $B_0$ increase is comparable with the standard deviation of this parameter.

Before, discussing the causes of incompressibility, we would like to comment on the possible effects of non-hydrostatic effects in our experiments. It is known that 16:3:1 methanol-ethanol-water is not a hydrostatic pressure-medium beyond 11 GPa[13]. This typically induces peak broadening, which in our case can be noticed in Fig. 2. However, in previous experiments, it has been established that non-hydrostatic stresses does not influence, at least up to 50 GPa, the structural stability of isomorphic compounds.[4] In particular, we have here checked that the lattice-parameters obtained from multiple diffraction lines of the same compound agree within their experimental uncertainties. Therefore, the possibility of a pressure-induced tetragonal or rhombohedral distortion of the cubic structure[16] caused by non-hydrostatic stresses can be excluded.

As described, the four studied materials are found to be quite incompressible. In particular, they are less compressible than $MoN$[18], cubic $Hf_3N_4$,[19] and cubic $Si_3N_4$[1] and as incompressible as cubic BN,[6] showing a bulk modulus only 16% smaller than diamond. The incompressible and non-deformable character of the studied compounds is related with the common features of their crystalline structures.[20] In all bimetallic nitrides (carbides), the Mo-N (Mo-C) bond distance is pretty similar to the Mo-N distance in rock-salt MoN (2.095 Å)[21]; e.g. Mo-N is 2.079(7) Å in $Ni_2Mo_3C_{0.52}N_{0.48}$. However, in our compounds, as a distinctive feature, the $NMo_6$ octahedra not only form a three-dimensional network, but additionally are interpenetrated by $NPd_5$ ($NNi_5$) trigonal dipyramids or $NFe_4$ ($NCo_4$) tetrahedra forming a very strong three-dimensional network, which give them their particular incompressibility. In particular, from our structural refinements we concluded that the Mo-N bond is quite incompressible, making consequently the bimetallic nitrides ultra-incompressible. The estimated bond compressibility in $Ni_2Mo_3C_{0.52}N_{0.48}$ is $\chi_{Mo-N} = -\partial \ln(d)/\partial P = 7.5(5)\ 10^{-4}$ $GPa^{-1}$, where $d$ represents the Mo-N bond distance. This bond compressibility is smaller than that observed for the uncompressible Mo-O bonds in several molybdates,[22-24] supporting our hypothesis. Regarding the Mo-C bond, we obtained from the structural refinements of $Fe_3Mo_3C$ that $\chi_{Mo-C} = 6.4(4)\ 10^{-4}$ $GPa^{-1}$. Thus, apparently this bond is even less compressible than the Mo-N bond, which is consistent with the fact that the substitution of N by C may produce a decrease of compressibility in the studied materials. This observation is also consistent with conclusions previously obtained for noble metal nitrides and carbides.[25]

In the past, based upon the calculated shear modulus ($G$),[4] interstitial nitrides have been considered as good candidates for hard materials. In the present case the shear modulus can be estimated using the determined bulk modulus. For an isotropic system $G$ can be directly related to $B_0$ according to $G = \frac{3}{2}B_0(1-2\nu)/(1+\nu)$ where $\nu$ is the Poisson's ratio, the value of the ratio of the transverse strain to the corresponding axial strain. For the compounds here studied a good approximation is to consider $\nu = 0.3$.[4] Under such approximation, for $Fe_3Mo_3C$ it is obtained $G = 172$ GPa. Then, employing the correlation between shear modulus and the Vickers hardness reported by Teter,[26] a hardness of 23.3 GPa is estimated for $Fe_3Mo_3C$. These value is larger than the Vicker hardness of cubic $Si_3N_4$,[1] thus pointing to a larger resistance of $Fe_3Mo_3C$ to plastic deformations. Therefore, the compounds here studied can be as hard as superhard nitrides. However, they are in the low hardness boundary for superhard materials.

In Summary, we have studied experimentally the compressibility and structural properties of bimetallic nitrides, carbides, and carbonitrides. They are stable in their cubic structure under extreme compression and have large bulk moduli. In particular $Fe_3Mo_3C$ has a bulk modulus comparable to cubic BN. We also found evidence suggesting that the substitution of N by C reduces the compressibility of the bimetallic family studied. The reported incompressibility is a consequence of the framework of short and strong Mo-N (Mo-C) bonds of the materials. The known correlation between shear modulus and hardness suggests that refractory interstitial nitrides and carbides are good candidates for ultra-hard materials. A Vickers hardness of 23.3 GPa is estimated for $Fe_3Mo_3C$. A possibility to explore in the search of reducing compressibility is the substitution of Mo by W, which is known to enhance incompressibility in transition-metal nitrides.[27] An advantage of interstitial nitrides and carbides is that they can be synthesized in large amounts without the use of high-pressure conditions. Another interesting fact is their peculiar magnetic properties, caused by the narrow *d*-band they have sitting at around the Fermi level.[28] We hope that this work will stimulate further research on this interesting family of compounds.

**Acknowledgments**

Research supported by Spanish MCYT under Grants No. MAT2010-21270-C04-01, MAT2009-14144-CO3-03, and CSD2007-00045 (MALTA Consolider Team). XRD experiments carried out at the Diamond Light Source (I15 beamline, proposal EE6517).

**References**

[1] A. Zerr, R. Riedel, T. Sekine, J.E. Lowther, W.Y. Ching, and I. Tanaka, Adv. Mater. **18**, 2933 (2006).

[2] A. Friedrich, B. Winkler, L. Bayarjargal, W. Morgenroth, E.A. Juarez-Arellano, V. Milman, K. Refson, M. Kunz, and K. Chen, Phys. Rev. Let. **105**, 085504 (2010).

[3] A. El-Himri, F. Sapiña, R. Ibañez, and J. Beltran, J. Mater. Chem. **11**, 2311 (2001).

[4] D. Errandonea, C. Ferrer-Roca, D. Martinez-Garcia, A.Segura, O. Gomis, A. Muñoz, P. Rodriguez-Hernandez, J. Lopez-Solano, S. Alconchel, and F. Sapiña, Phys. Rev. B **82**, 174105 (2010).

[5] S. Alconchel, F. Sapiña, and E. Martinez, Dalton Trans. 2463 (2004).

[6] V.V. Brazhkin, A.G. Lyapin, and R.J. Hemley, Philos. Mag. A **82**, 231 (2002).

[7] S. Alconchel, F. Sapiña, D. Beltrán, and A. Beltrán, J. Mater. Chem. **8**, 1901 (1998).

[8] S. Alconchel, F. Sapiña, D. Beltrán, and A. Beltrán, J. Mater. Chem. **9**, 749 (1999)

[9] S. Alconchel, B. Pierini, F. Sapiña, and E. Martinez, Dalton Trans. 330 (2009).

[10] H.K. Mao, J. Xu, and P.M. Bell, J. Geophys. Res. **91**, 4673 (1986).

[11] M. Yokoo, N. Kawai, K.G. Nakamura, K. Kondo, Y. Tange, and T. Tsuchiya, Phys. Rev. B **80**, 104114 (2009).

[12] D. Errandonea, Y. Meng, M. Somayazulu, and D. Häusermann, Physica B **355**, 116 (2005).

[13] S. Klotz, J.C. Chervin, P. Munsch, and G. Le Marchand, J. Phys. D: Appl. Phys. **42**, 075413 (2009).

[14] D. He and T.S. Duffy, Phys. Rev. B **73**, 134106 (2006).

[15] D. Errandonea, R. Boehler, S. Japel, M. Mezouar, and L.R. Benedetti, Phys. Rev. B **73**, 092106 (2006).

[16] D. Errandonea, R.S. Kumar, F.J. Manjon, V.V. Ursaki, and E.V. Rusu, Phys. Rev. B **79**, 024103 (2009).

[17] A.H. Reshak, S. Auluck, and I.V. Kityk, J. Phys. Chem. B **115**, 3363 (2011).

[18] E. Zhao, J. Wang, and Z. Wu, Phys. Stat. Sol. B **247**, 1207 (2010).

[19] M. Mattesini, R. Ahuja, and B. Johansson, Phys. Rev. B **68**, 184108 (2003).

[20] K.S. Weil, P.N. Kunta, and J. Grins, J. Solid State Chem. **146**, 22 (1999).

[21] F. Fujimoto, Y. Yakane, M. Satou, F. Komori, K. Agate, and Y. Andob, Nucl. Instrum. Methods Phys. Res. B **19**, 791 (1987).

[22] D Errandonea, D. Santamaria-Perez, S. N. Achary, A. K. Tyagi, P. Gall, and P. Gougeon, J. Appl. Phys. **109**, 043510 (2011).

[23] D. Errandonea, D. Santamaria-Perez, V. Grover, S. N. Achary, and A. K. Tyagi, J. Appl. Phys. **108**, 073518 (2010).

[24] D. Errandonea, R. S. Kumar, X. H. Ma, and C. Y. Tu, J. Solid State Chem. **181** 355 (2008).

[25] C.-Z. Fan, S.-Y. Zeng, Z.-J. Zhan, R.-P. Liu, W.-K. Wang, P. Zhang, and Y.-G. Yao, Appl. Phys. Letters **89**, 071913 (2006).

[26] D. M. Teter, MRS Bull. **23**, 22 (1998).

[27] Y. Benhoi, W. Chuenlei, S. Xuenyu, S. Qiuju, and C. Dong,. J. Alloys Compds. **487**, 556 (2009).

[28] T. Waki, S. Terazawa, Y. Umemoto, Y. Tabata, K. Sato, A. Kondo, K. Kindo, and H. Nakamura, J. Phys: Conf. Series 320, 012069 (2011).

**Table I.** EOS parameters and unit-cell parameter (*a*) determined at ambient pressure.

| Compound | *a* (Å) | $V_0$ (Å$^3$) | $B_0$ (GPa) | $B_0'$ |
|---|---|---|---|---|
| $Pd_2Mo_3N$ | 6.817(5) | 316.9(2) | 333(3) | 4.3(2) |
| $Ni_2Mo_3C_{0.52}N_{0.48}$ | 6.645(5) | 293.4(2) | 344(4) | 3.8(2) |
| $Co_3Mo_3C_{0.62}N_{0.38}$ | 11.069(8) | 1356(0.5) | 353(4) | 4.7(3) |
| $Fe_3Mo_3C$ | 11.099(8) | 1368(0.5) | 374(3) | 3.7(3) |

**Figure captions**

**Figure 1:** (color online) Schematic view of the crystal structure of $Pd_2Mo_3N$ and $Fe_3Mo_3C$.

**Figure 2:** (color online) ADXRD data of $Ni_2Mo_3C_{0.52}N_{0.48}$ and $Fe_3Mo_3C$ at selected pressures and RT. Measured (dots) and calculated (solid lines) patterns are shown together with the difference curve (dashed lines) and calculated positions of Bragg reflections (ticks).

**Figure 3:** Unit-cell volume versus pressure for studied compounds. Experimental data: Solid (empty) symbol ruby (gold) scale. Lines: reported EOS. For $Fe_3Mo_3C$ and $Co_3Mo_3C_{0.62}N_{0.38}$ we plotted V/4 to facilitate comparison with $Pd_2Mo_3N$, and $Ni_2Mo_3C_{0.52}N_{0.48}$.

**Figure 1**

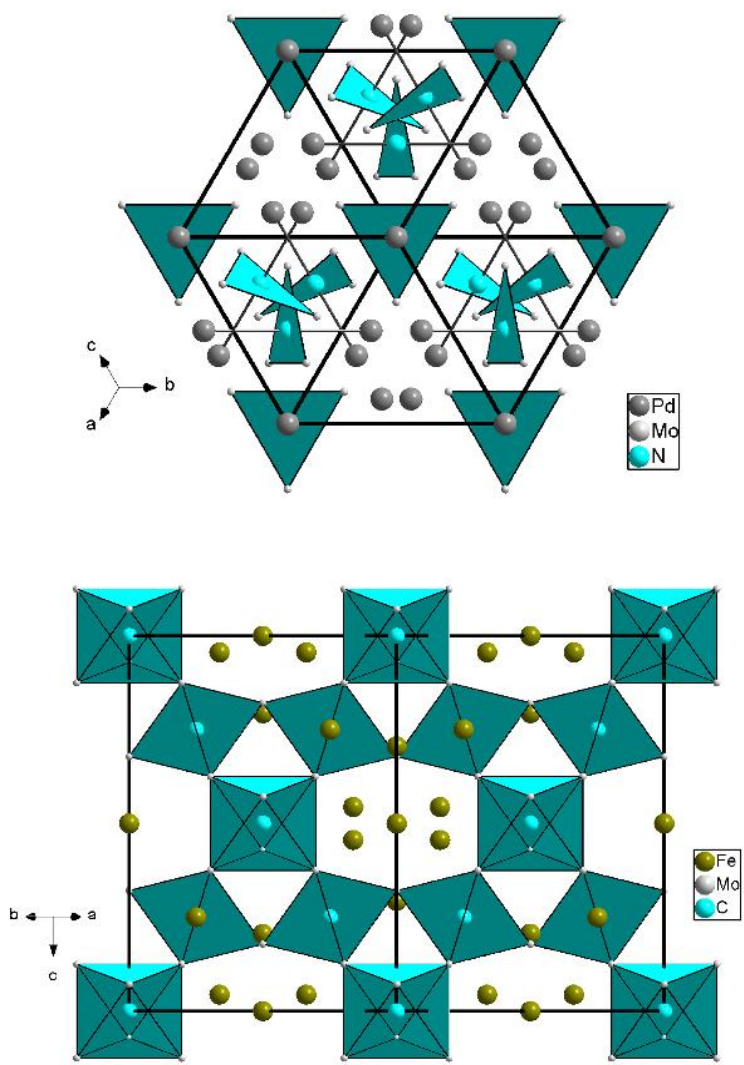

**Figure 2**

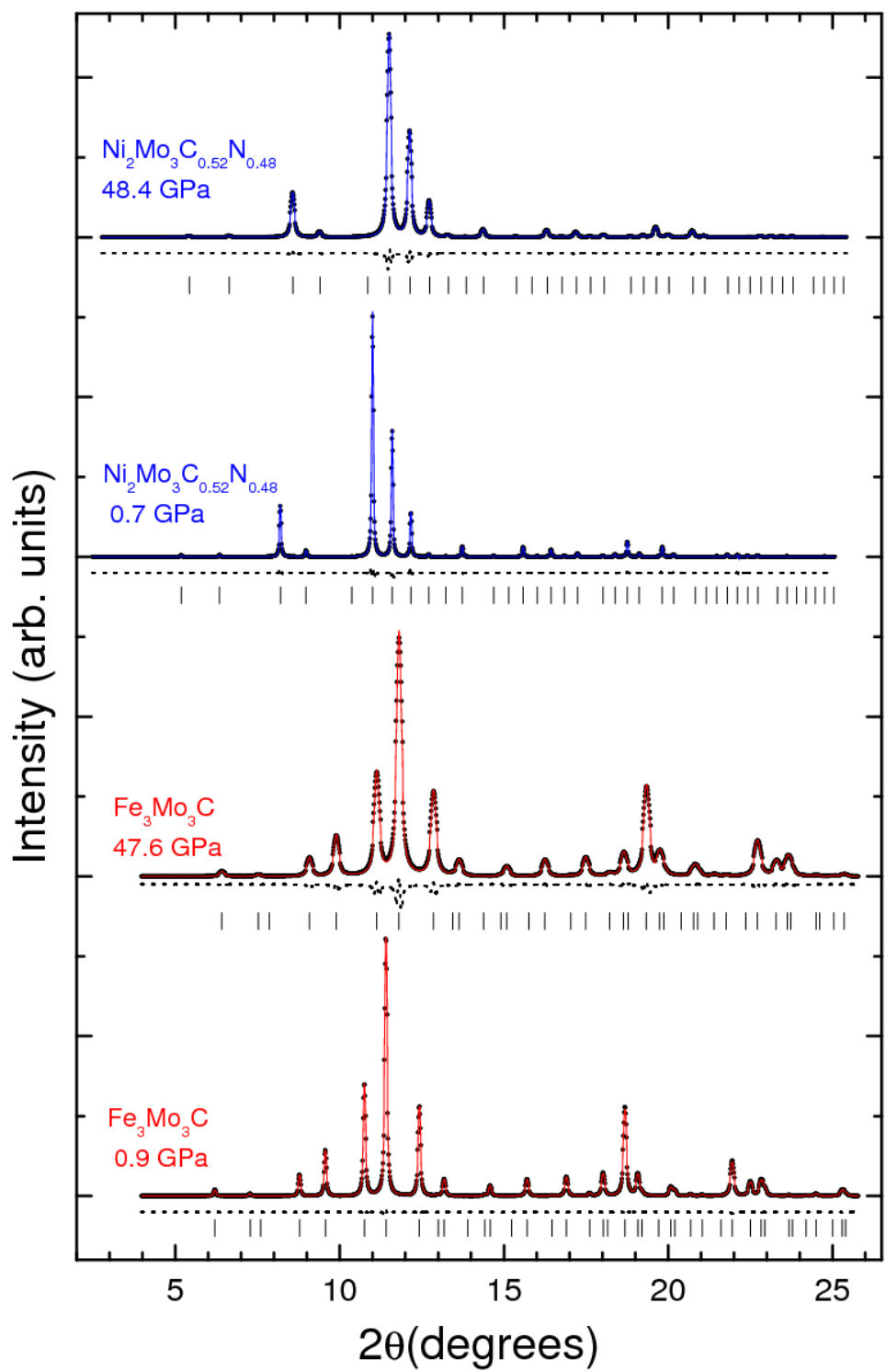

**Figure 3**

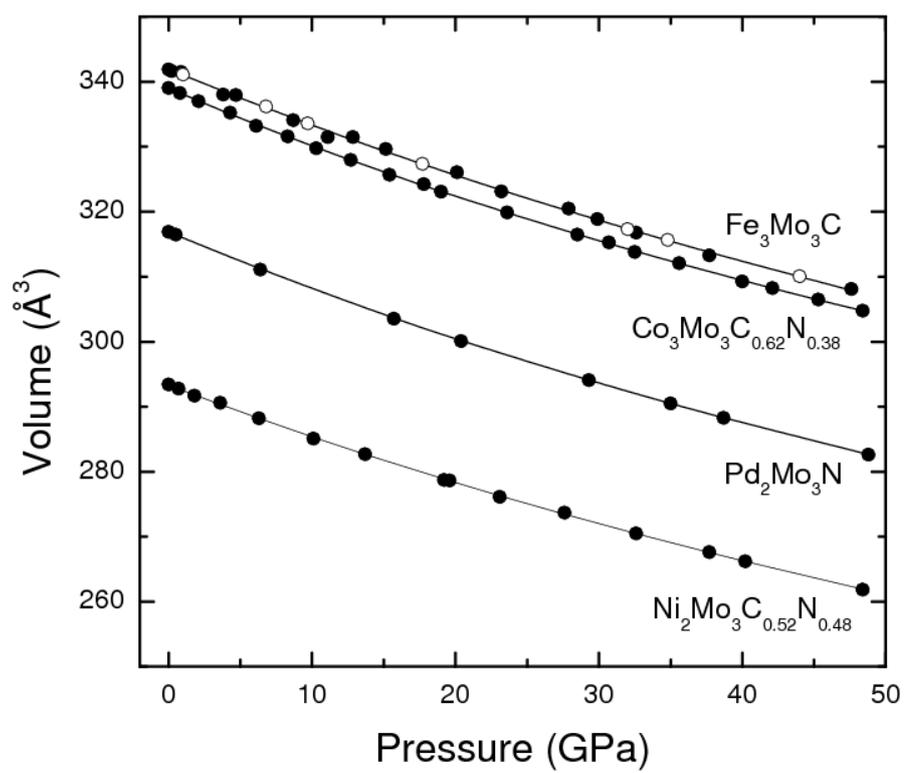